\documentclass[aip,cha,amsmath,amssymb,reprint]{revtex4-2}

\usepackage{graphicx}
\usepackage{dcolumn}
\usepackage{bm}
\usepackage{subfigure}
\usepackage{multirow}
\usepackage[utf8]{inputenc}
\usepackage[T1]{fontenc}
\usepackage{mathptmx}
\usepackage{etoolbox}
\usepackage{lipsum}
\usepackage{xcolor}
\usepackage{xspace, units}
\usepackage{soul}

\DeclareMathAlphabet{\pazocal}{OMS}{zplm}{m}{n}

\newcommand {\CC}{\ensuremath{\pazocal{C}}}

\newcommand{\I}{\ensuremath\pazocal{I}}

\newcommand{\A}{\ensuremath\pazocal{A}}
\newcommand{\B}{\ensuremath\pazocal{B}}

\newcommand{\revise}[1]{{\color{black}#1}}

\newcommand{\halpha}{\hat{\alpha}}
\newcommand{\hA}{\hat{A}}
\newcommand{\hB}{\hat{B}}

\newcommand{\hbeta}{\hat{\beta}}
\newcommand{\hgamma}{\hat{\gamma}}
\newcommand{\htau}{\hat{\tau}}

\newcommand{\djsc}{D_{\textnormal{JS}}^C}
\newcommand{\tmi}{T^{(\htau)}}

\makeatletter
\def\@email#1#2{%
 \endgroup
 \patchcmd{\titleblock@produce}
  {\frontmatter@RRAPformat}
  {\frontmatter@RRAPformat{\produce@RRAP{*#1\href{mailto:#2}{#2}}}\frontmatter@RRAPformat}
  {}{}
}%
\makeatother

\usepackage{mathtools}

\DeclarePairedDelimiterX{\infdivx}[2]{(}{)}{%
	#1\;\delimsize\|\;#2%
}
\newcommand{\infdiv}{E_\text{KL}\infdivx}
\newcommand{\infdivD}{D^C_\textnormal{JS}\infdivx}

\DeclareFontFamily{U}{matha}{\hyphenchar\font45}
\DeclareFontShape{U}{matha}{m}{n}{
	<5> <6> <7> <8> <9> <10> gen * matha
	<10.95> matha10 <12> <14.4> <17.28> <20.74> <24.88> matha12
}{}
\DeclareSymbolFont{matha}{U}{matha}{m}{n}

\DeclareMathSymbol{\Lt}{3}{matha}{"CE}

\date{\today}

\begin{document}
\title{Transcript-based estimators for characterizing interactions}
\author{Manuel Adams}
\email{ma@uni-bonn.de}
\affiliation{Department of Epileptology, University of Bonn Medical Centre, Venusberg Campus 1, 53127 Bonn, Germany}
\author{Jos\'{e} M. Amig\'{o}}
\email{jm.amigo@umh.es}
\affiliation{Centro de Investigaci\'{o}n Operativa, Universidad Miguel Hern\'{a}ndez, 03202 Elche, Spain}
\author{Klaus Lehnertz}
\email{klaus.lehnertz@ukbonn.de}
\affiliation{Department of Epileptology, University of Bonn Medical Centre, Venusberg Campus 1, 53127 Bonn, Germany}
\affiliation{Helmholtz Institute for Radiation and Nuclear Physics, University of Bonn, Nussallee 14--16, 53115 Bonn, Germany}
\affiliation{Interdisciplinary Center for Complex Systems, University of Bonn, Br{\"u}hler Stra\ss{}e 7, 53175 Bonn, Germany}

\begin{abstract}
    The concept of transcripts was introduced in 2009 as a means to characterize various aspects of the functional relationship between time series of interacting systems.
    \revise{Based on this concept that utilizes algebraic relations between ordinal patterns derived from time series, estimators for the strength, direction, and complexity of interactions have been introduced. These 
    estimators, however, have not yet found widespread application
    in studies of interactions between real-world systems.}
    Here, we revisit the concept of transcripts and \revise{showcase the usage of transcript-based estimators for a time-series-based investigation of interactions between coupled paradigmatic dynamical systems of varying complexity.}
    At the example of a time-resolved analysis \revise{of multichannel and multiday recordings of ongoing human brain dynamics, we demonstrate the potential of the methods to provide novel insights into the intricate spatial-temporal interactions in the human brain underlying different vigilance states.}
\end{abstract}
\maketitle

\begin{quotation}
Synchronization is a widespread natural phenomenon that requires suitable and robust time-series-analysis approaches to decipher the multifaceted nature of interactions.
\revise{Algebraic relations between ordinal patterns --~or transcripts~--} derived from pairs of time series can be utilized to characterize the complexity, strength, and direction of an interaction.
We revisit this approach and demonstrate its suitability for providing relevant insights into how interactions in the human brain are being shaped during wakefulness and sleep.
\end{quotation}
\frenchspacing
\section{Introduction}
Many natural and man-made systems are composed of a large number of interacting subsystems.
Understanding how interactions between parts of a system shape the dynamics of a system as a whole is of great interest in diverse scientific fields, ranging from physics via earth science and climatology to economics and neuroscience~\cite{blekhman1988,pikovsky2001,boccaletti2002,kantz2003,strogatz2004,marwan2007,hlavackova2007,arenas2008,Ghosh2022,varela2001,pereda2005,fell-axmacher2011}.
Direct access to interactions and their time-dependencies is often not possible, so linear and non-linear bivariate time-series-analysis techniques are \revise{frequently} used to quantify properties of an interaction from a pair of time series of appropriate system observables.
Given that interactions can manifest themselves in various aspects of the dynamics, analysis techniques focus on quantifying the strength and/or direction of an interaction and, in the best case, on functional relationships. 
Many of these analysis techniques, originating from statistics, synchronization theory, non-linear dynamics, information theory, statistical physics, and from the theory of stochastic processes, yield promising results.
Among information-theory-based analysis techniques, particularly approaches building on ordinal patterns~\cite{bandtpompe2002,Amigo2008} have received much attention, since they constitute an efficient and versatile methodology for characterizing interactions~\cite{Staniek2008,echegoyen2019,Leyva2022,YamashitaRiosdeSousa2022,amigo2023ordinal,Lehnertz2023}. 

The concept of transcripts exploits how a sequence of ordinal patterns representing one time series maps onto the sequence from another time series.
\revise{To this end, algebraic relations between ordinal patterns are exploited that}
can be efficiently determined in a permutation-based manner and can be utilized to characterize the strength, direction, and complexity of an interaction~\cite{Monetti2009,amigo2012,Monetti2013CC,Monetti2013TMI}.
Despite the simplicity of this concept, transcripts have mainly been utilized to characterize different states of synchronization in coupled model systems, and little is known about their usefulness for characterizing interactions in real-world systems based on empirical data.

Here, we revisit the concept of transcripts and review the properties of transcript-based  estimators for various properties of an interaction. 
With an investigation of multichannel, multiday electroencephalographic recordings from healthy subjects, we demonstrate the suitability of a transcript-based characterization to provide relevant insights into brain-wide interactions during wakefulness and sleep.
%
\section{Methods}
\revise{In the following, we consider two time series, each derived from a suitable observable
of two (supposedly coupled) dynamical systems $X$ and $Y$.
Time series have a length $N$.\\
}

\subsection{\revise{Ordinal patterns and transcripts}}
\begin{figure*}[htbp]
   \centering
    \includegraphics[width=\linewidth]{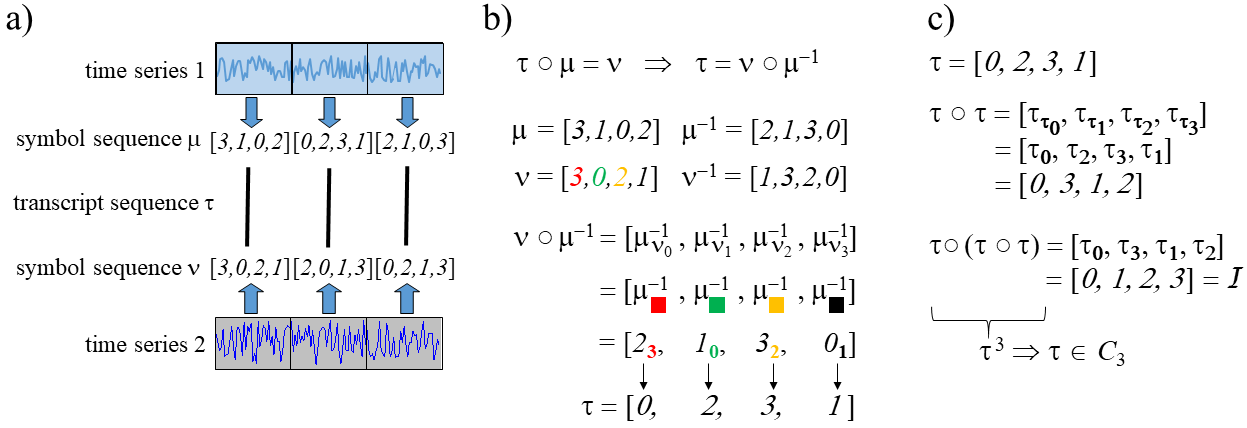}
    \caption{\revise{
    Illustrative example for calculating the transcript $\tau$ between two time-series-derived ordinal patterns $\mu$ and $\nu$ as well as the order class $C_n$ to which $\tau$ belongs. 
		\textbf{a}) Schematic of time-resolved derivation of ordinal-pattern (symbol) sequences ($d=4$) from two time series.  
		\textbf{b}) Exemplary calculation of transcript $\tau$ from the composition of ordinal pattern $\nu$ with the inverse of ordinal patterns $\mu$ denoted as $\mu^{-1}$ (cf. Eq.~\ref{eq:transcript}). 
		Colors highlight the entries of $\nu$ that are composed elementwise with $\mu^{-1}$ to obtain $\tau$. 
		For example, $\textit{2}_\textrm{\bf {\color{red} 3}}$ means: exchange entry $\textit{2}$ (at position 0 in $\mu^{-1}$) with the entry at position 3, i.e., $\textit{0}$. Note that the  composition $\mu$ with $\nu^{-1}$ can yield a different transcript $\tau$.
		\textbf{c}) Exemplary derivation of an order class. 
        The transcript $\tau = \left[\textit{0,2,3,1}\right]$ needs to be composed with itself three times to achieve the unit symbol $\I$. 
		This transcript thus belongs to order class $C_3$.
    }
    }
    \label{fig:schematic}
\end{figure*}

\revise{In order to derive a sequence of ordinal patterns from a time series, we divide the latter into consecutive delay vectors of size $d$ using a delay embedding~\cite{takens1981,bandtpompe2002,staniek2007} with delay $m$.
This then leads to a sequence of $L=N-(d-1)m$ delay vectors \revise{(cf. Fig.~\ref{fig:schematic}{\bf a})} and for each delay vector, we derive an ordinal pattern or symbol by rank-ordering the amplitude values from lowest to highest, thus creating values between 0 and $d-1$.}
For example, with $d=4$ and $m=1$, for the delay vector $q=(1.7,1.2,1.3,1.5)$ the corresponding ordinal pattern is $\phi=[\phi_0,\phi_1,\phi_2,\phi_3]=[1,2,3,0]$.
\revise{In case a delay vector contains the same value multiple times, we order the entries from left to right, i.e., the unit symbol is $\I=[0,1,2,\ldots,d-1]$}. 
The latter thus relates to both, a delay vector containing constant or monotonously increasing values.\\ 

\revise{Given two ordinal patterns $\mu$ and $\nu$ from the set $S_d$ of ordinal patterns of length $d$, we can use the framework of permutation theory to derive a means to characterize \revise{algebraic relations} between $\mu$ and $\nu$.} 
\revise{The transcript $\tau \in S_d$ with elements $[\tau_0,\tau_1,\ldots,\tau_{d-1}]$ from $\mu=[\mu_0,\mu_1,\ldots,\mu_{d-1}]$ to $\nu=[\nu_0,\nu_1,\ldots,\nu_{d-1}]$ is defined as 
\begin{equation} \label{eq:transcript}
		\tau \circ \mu = \nu,
\end{equation}
where each element of $\mu, \nu,$ and $\tau$ is an element of the set $\{0,1,\ldots,d-1\}$. 
The two ordinal patterns $\tau $ and $\mu $ act onto each other via composition, so that Eq.~\ref{eq:transcript} becomes}
\begin{equation}
		\tau \circ \mu = [\mu_{\tau_0},\mu_{\tau_1},\ldots,\mu_{\tau_{d-1}}]=[\nu_0,\nu_1,\ldots,\nu_{d-1}].
\end{equation}

\revise{One can find the inverse transcript $\tau^{-1}$  by the following considerations (we will omit the composition symbol $\circ$ in the following). 
Since we know that
\begin{equation}
		\tau^{-1}\tau = [\tau_{(\tau^{-1})_0},\tau_{(\tau^{-1})_1},\ldots,\tau_{(\tau^{-1})_{d-1}}]\stackrel{!}{=}\I,
\end{equation}
it follows directly that 
\begin{equation}
		\tau_{(\tau^{-1})_0}<\tau_{(\tau^{-1})_1}<\ldots<\tau_{(\tau^{-1})_{d-1}},
\end{equation}
which incidentally shows that $\tau^{-1}$ is the ordinal pattern of the delay vector $(\tau _{0},\tau_{1},...,\tau _{d-1})$.
With the above definitions, a transcript --~being itself an ordinal pattern~-- can thus be calculated via $\tau = \nu\mu^{-1}$ \revise{(cf. Fig.~\ref{fig:schematic}{\bf b})}}. \\
	
\subsection{Transcript-based estimators for complexity, strength, and direction of an interaction}
\revise{Estimators for the complexity of an interaction can be derived from sequences of both, transcripts and ordinal patterns.\\

Different transcripts indicate different algebraic relations between ordinal patterns, and the complexity of a transcript $\tau$ can be assessed from its order $n$~\cite{Monetti2009}.
The latter describes the number of non-trivial operations needed to obtain the unit symbol (cf. Fig.~\ref{fig:schematic}{\bf c}),
\begin{equation}
        n=\underset{n\in\mathbb{N}_{>0}}{\arg\,\min}(\tau^n=\I).
\end{equation}
Since the unit symbol, when occurring in a sequence of transcripts, denotes a relation of minimal Kolmogorov complexity~\cite{Kolmogorov1998} between ordinal patterns, the order $n$ can be interpreted as describing the complexity of the relations by means of the number of non-trivial operations needed to return to the relation of the lowest Kolmogorov complexity (described by $\I$).  

Transcripts with order $n$ can be assigned to order class $C_n\in \mathbb{N}_{>0}$, which is a subset of the group of ordinal patterns ($S_d=\cup\, C_n$).
Estimating the probability densities of different order classes enables to characterize the Kolmogorov complexity of a sequence of transcripts and thus the complexity of a relations between two time series~\cite{Monetti2009}, thereby potentially
providing complementary information about an interaction. 
The expected values for the respective probabilities of each order class can be derived analytically~\cite{Wilf1986}. \\
}
 
\revise{
We proceed with an estimator for the complexity of an interaction that is derived from two sequences of ordinal pattern $\halpha$ and $\hbeta$ with their respective entries $\{\halpha_i\}_{i=0}^{L-1}$ and $\{\hbeta_i\}_{i=0}^{L-1}$ alongside with the sequence of transcripts $\htau^{(\halpha,\hbeta)}$ with $\htau^{(\halpha,\hbeta)}_i \halpha_i = \hbeta_i$. 
With the entropies of the sequences $H(\halpha)$, $H(\hbeta)$, and $H(\htau^{(\halpha,\hbeta)})$ together with the entropy of the joint ordinal pattern sequences $H(\halpha,\hbeta)$,
complexity $\CC$ of an interaction (or coupling complexity) can be defined as~\cite{amigo2012,Monetti2013CC} 
\begin{equation}\label{eq:cc}
		\CC(\halpha,\hbeta) = \min[H(\halpha),H(\hbeta)]-(H(\halpha,\hbeta)-H(\htau^{(\halpha,\hbeta)})).
\end{equation}
Note that $\CC$ characterizes the relation between two time series~\cite{amigo2012}, adjusted by the entropy of the transcription process that captures the relation between the two time series from a different perspective.
$\CC$ is a symmetric measure under the interchange of sequences $\halpha$ and $\hbeta$
and is bounded between 0 and $\CC=\min\left[H(\halpha),H(\hbeta)\right]=\log d! -(2\log 2 / d!)$\;\cite{Monetti2013CC}. 
The upper bound quantifies the maximum information that can be gained about an interaction.
The lower bound ($\CC=0$) indicates both, the case of comparing two arbitrary time series that yield completely equal distributions of ordinal patterns (which can occur for completely synchronized systems) and the case of comparing an arbitrary time series to a completely random one}~\cite{Monetti2013CC}. 
In both cases, no information is gained about the interaction between the two \revise{systems} by investigating the respective distributions of ordinal patterns.\\
	
\revise{It can be expected that the weaker (stronger) the interaction between two systems the more dissimilar (similar) are their dynamics (or properties thereof). 
As an estimator of the strength of an interaction, one might therefore consider an entropy-based measure of dissimilarity between two probability density functions. 
Monetti and colleagues~\cite{Monetti2009} proposed to employ the Kullback-Leibler divergence to compare the probability densities of the order classes of coupled systems with the probability densities that are expected for independent systems.
Here, we compute the Jensen–Shannon (JS) divergence, which is itself a symmetric and bounded distance measure derived from the Kullback–Leibler divergence~\cite{topsoe2002}. 
}    
\revise{To begin with, let us}
define the probability densities for transcripts as
\revise{
\begin{equation}
		P(\tau) = \sum_{\mu, \nu \in S_d: \nu\mu^{-1}=\tau}P^J(\mu,\nu),
\end{equation}
}
with $P^J$ denoting the joint probability of the pair \revise{$\mu$, $\nu$} in ordinal patterns sequences $\halpha$ and $\hbeta$, and 
\revise{
\begin{equation}
		P^\text{ind}(\tau) = \sum_{\mu, \nu \in S_d: \nu\mu^{-1}=\tau}P_{\halpha}(\mu)P_{\hbeta}(\nu)
\end{equation}
}
the probability density of transcripts for independent pairs. 
Here, $P_{\halpha}(\mu)$ ($P_{\hbeta}(\nu)$) denotes the probability of ordinal pattern $\mu$ ($\nu$) in the ordinal pattern sequences $\halpha$ ($\hbeta$).
By the same token, 
\revise{one can 
define probability densities
for order classes $C_n$ by restricting the sums over ordinal patterns to sums over order classes:} 
\begin{equation}\label{eq:pj}
		P_{C_n}\coloneqq \sum_{\tau \in C_n}P(\tau),
	\end{equation}
	\begin{equation}\label{eq:pind}
		P_{C_n}^{\text{ind}}\coloneqq \sum_{\tau \in C_n}P^\text{ind}(\tau).
\end{equation}
\revise{
With the mixing probability
\begin{equation}\label{eq:pm}
		P_{C_n}^{\text{M}}\coloneqq \frac{1}{2}P_{C_n}+\frac{1}{2}P_{C_n}^{\text{ind}},
\end{equation}
}
\revise{we now make use of the Kullback-Leibler divergence for order classes, which for two arbitrary distributions of order classes $P_{C_n}^{(1)}$ and $P_{C_n}^{(2)}$ can be defined as~\cite{Monetti2009}:
\begin{equation}\label{eq:ekl}
		\infdiv{P^{(1)}_C}{P^{(2)}_C}=\sum_{n}P^{(1)}_{C_n}\log_2\left(P^{(1)}_{C_n}/P^{(2)}_{C_n}\right).
\end{equation}
}
\revise{The above definitions can then be used to derive the JS-divergence for order classes as}%
\begin{equation}\label{eq:djsc}
		\infdivD{P_{C}}{P_{C}^\text{ind}}= \frac{1}{2} (\infdiv{P_C}{P_C^{M}}+ \infdiv{P_C^{\text{ind}}}{P_C^{M}}).
\end{equation}
\revise{Note that since $E_\text{KL}\geq 0$, we have $\djsc\geq 0$. 
Its upper bound depends on the embedding dimension $d$ used to derive ordinal pattern and can be determined analytically (Appendix \ref{app:maxdjsc}).}\\
	
\revise{Finally, we recall the definition of a transcript-based estimator for the direction of an interaction~\cite{Monetti2013TMI}, which builds on a time-delayed mutual information~\cite{fraser1986,kaneko1986,Schreiber2000}. 
For three arbitrary ordinal pattern sequences $\halpha$, $\hbeta$ and $\hgamma$, it is defined as
\begin{equation}
		I(\htau^{(\hgamma,\halpha)},\htau^{(\hbeta,\halpha)}) = H(\htau^{(\hgamma,\halpha)}) - H(\htau^{(\hgamma,\halpha)}|\htau^{(\hbeta,\halpha)}),
		\label{meth:tmi}
\end{equation}
with $\htau^{(\halpha,\hbeta)}$ denoting the sequence of transcripts between the sequences of ordinal patterns $\halpha$ and $\hbeta$.
In order to introduce a delay, 
we use an approach similar to the one employed for the definition of the symbolic transfer entropy~\cite{Staniek2008,dickten2014}, and define}
\begin{equation}
		\begin{aligned} 
			\htau^{(\halpha,\halpha^*)}\halpha &= \halpha^*\;, \\ 
			\htau^{(\hbeta,\hbeta^*)}\hbeta &= \hbeta^*\;,
		\end{aligned} \qquad
		\begin{aligned}
			\halpha^*_i&=\halpha_{i+\Lambda} \;,\\ 
			\hbeta^*_i &=\hbeta_{i+\Lambda} \;,
		\end{aligned}
		\label{meth:tmi2}
\end{equation}
where $\halpha^*$ denotes the ordinal pattern sequences $\halpha$ shifted elementwise in time by a delay $\Lambda$ (which is not to be confused with the embedding delay $m$). 
Using Eq.~(\ref{meth:tmi}) alongside with the asymmetry of the mutual information of transcripts under the interchange of the sequences $\halpha$ and $\hbeta$, the transcript-based directionality index can then be defined as
\revise{
\begin{equation}\label{eq:dirindex} 
		\tmi = I(\htau^{(\hbeta,\hbeta^*)},\htau^{(\halpha,\hbeta)})-I(\htau^{(\halpha,\halpha^*)},\htau^{(\hbeta,\halpha)}),
\end{equation}
with $T^{\htau}$ exploiting the asymmetry of Eq. (\ref{meth:tmi}) under exchange of $\halpha$ and $\hbeta$.

$\tmi$ takes on positive (negative) values, if system $X$ ($Y$) drives system $Y$ ($X$).
More precisely, we have 
\begin{equation}
    -\log d! \leq \tmi \leq \log d!,
\end{equation}
which can easily be deduced from the upper bound of the mutual information (Appendix \ref{app:boundtmi}). 
}
The case of $\tmi=0$ can refer to bidirectionally and uncoupled systems as well as for infinitely strong coupling \revise{(complete synchronization)}. In all these cases, there is no predominant flow of information \revise{and thus no indication of a direction of an interaction.}

\revise{
\section{Application to coupled model systems}
}
We begin by demonstrating the applicability of transcript-based \revise{estimators for properties of an interaction through an investigation of coupled model systems with well-known 
characteristics}.
The first system is exemplary for coupled discrete iterative maps and consists of two unidirectionally coupled Hénon maps\cite{Schiff1996}
\begin{equation}
	\begin{aligned}
		x_{1,i+1} &= a -x_{1,i}^2+b_1y_{1,i}, \\
		y_{1,i+1} &= x_{1,i}, \\
		x_{2,i+1} &= a - (kx_{1,i}x_{2,i}+(1-k)x_{2,i}^2) + b_2 y_{2,i}, \\
		y_{2,i+1} &= x_{2,i}.
	\end{aligned}
\end{equation}
We let $a=1.4$ and $b_1=b_2=0.3$ and vary the coupling strength $k$ from 0 to 1 in steps of $0.01$, with initial conditions drawn randomly from the interval [0,1]. After discarding transient data, we iterate the system for $N$ time steps.\\

The second system is exemplary for coupled oscillators and consists of a Rössler oscillator that is unidirectionally driven by another Rössler oscillator with mismatched control parameters
\begin{equation}\label{eq:roessler}
	\begin{aligned}
		\dot{x}_{1} &= -(y_1+z_1),\\
		\dot{y}_{1} &= x_1+0.2y_1,\\
		\dot{z}_{1} &= 0.2+z_1(x_1-5.7), \\
		\dot{x}_{2} &= -(y_2+z_2)+k(x_2-x_1),\\
		\dot{y}_{2} &= x_2+0.2y_2,\\
		\dot{z}_{2} &= 0.2+z_2(x_2-7). \\
	\end{aligned}
\end{equation}
This system exhibits transitions between different types of synchronization with increasing coupling strength, which we vary from $k=0$ to $k=1.5$ in steps of 0.001, increasing the step size to 0.005 for $1.5 < k \leq 3.0$ for the sake of computation time. 
We integrate the equations of motion using a fifth(fourth)-order Runge-Kutta integrator of the Dormand-Prince class, choosing an integration step $\delta t = 0.001$ and random initial conditions from the interval [-5,5]. 
After discarding transient data, we iterate the system for $N$ time steps and sample the data with a sampling interval $\Delta t = 0.01$.\\

\begin{figure}[htbp]
   \centering
    \includegraphics[width=\linewidth]{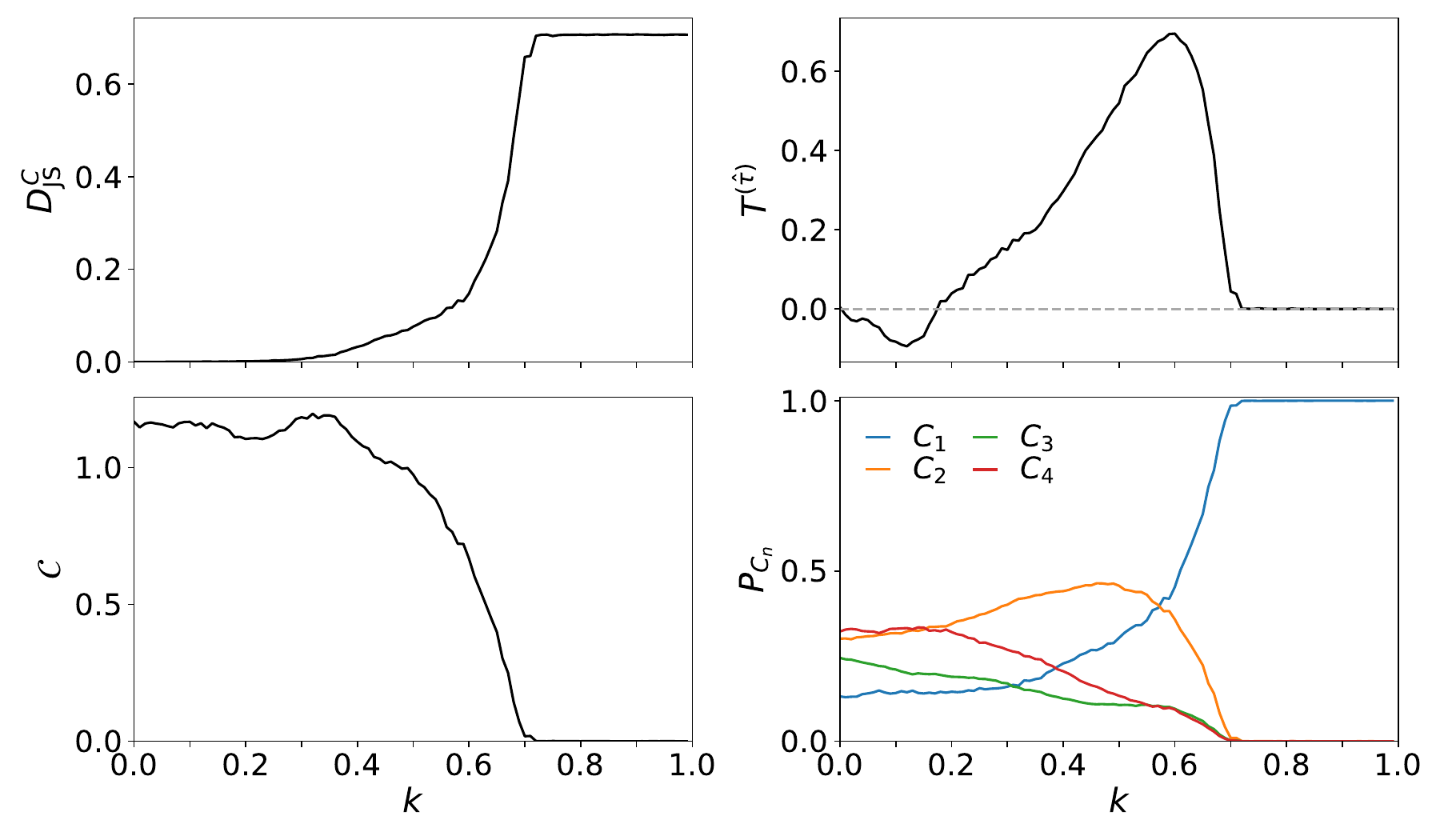}
    \caption{Dependence of estimators for the strength (top left), directionality (top right) and complexity (bottom left) of interaction on coupling strength $k$ for unidirectionally coupled Hénon maps. The dependence of the probability densities of order classes on coupling strength is shown in bottom right part of the figure.
    We generated time series of length $N=2^{17}$ of the $x$-components and derived ordinal patterns using an embedding dimension $d=4$ and an embedding delay $m=1$ (first zero-crossing of the autocorrelation function).
    }
    \label{fig:measureshenon}
\end{figure}

Figure~\ref{fig:measureshenon} summarizes our findings for the unidirectionally coupled Hénon maps.
The Jensen-Shannon divergence for order classes remains close to 0 up to a coupling strength of $k\approx 0.4$, indicating almost no interaction. For $k\geq0.4$, we observe a sudden increase until $\djsc$ reaches its maximum at $k\geq 0.7$, for which the \revise{maps} are fully coupled. \\
For low to intermediate coupling strengths $k\in [0,0.5]$, the transcript-based directionality index $\tmi$, (Eq. \ref{eq:dirindex}), 
indicates a direction of interaction from 
\revise{the second to the first map}, which is contrary to our construction of the system. 
However, since $D_{JS}^C$ indicates almost no interaction in this coupling range, the observed indication for directionality at lower coupling strengths is questionable~\cite{lehnertz-dickten2015}.
For larger coupling strengths, $\tmi>0$ up to  \revise{$k=0.7$},  while $\tmi=0$ for $k>0.7$, since no information flow is present anymore.\\
In terms of the probability densities of the order classes, for weak coupling ranges $k\in[0,0.2]$, the transcripts mainly
\revise{belong to order class}
$C_4$, indicating a higher complexity of the interaction.
As the coupling strength increases, we observe an increase in $P_{C_2}$, with $P_{C_3}$ and $P_{C_4}$ decreasing. For $k\approx 0.5$, $P_{C_2}$ reaches a maximum, with $C_2$ becoming the predominant order class.
For higher coupling strengths, $P_{C_1}$ increases strongly, with the \revise{unit symbol} $\I\in C_1$ being the only transcript left for $k \geq 0.7$, for which the coupling forces the \revise{maps} to fully synchronize.
This indicates that as the coupling strength is increased, the complexity of the interaction decreases as the \revise{individual dynamics} become more and more similar.\\ 
Coupling complexity $\CC$ corroborates this notion in part, fluctuating around $\CC\approx1.2$ for low to intermediate coupling strengths $k \in [0,0.3]$. 
The only difference here is that the highest value of $\CC$ is reached for $k\approx0.35$, at the crossing point of $P_{C_1}$ and $P_{C_3}$. For higher coupling strengths $k>0.35$, $\CC$ decreases with increasing similarity of the \revise{individual dynamics}
until it reaches 0 for $k=0.7$ where the dynamics become identical. 
We can observe here that coupling complexity seemingly becomes larger for coupling strengths, at which the densities of order classes lie closer to each other. 
The fluctuations that can be observed for $P_{C_n}$ and $\CC$ indicate that the complexity estimators may be sensitive to properties of the dynamics, which are not captured by the estimators for strength and direction of interactions. The results for $P_{C_n}$ and $\CC$ also indicate that interactions for different coupling strengths are unlikely to be described by a single function, \revise{but can instead be described by a unique functional relationship for each coupling strength.} \\

\begin{figure}[htbp]
    \centering    
    \includegraphics[width=\linewidth]{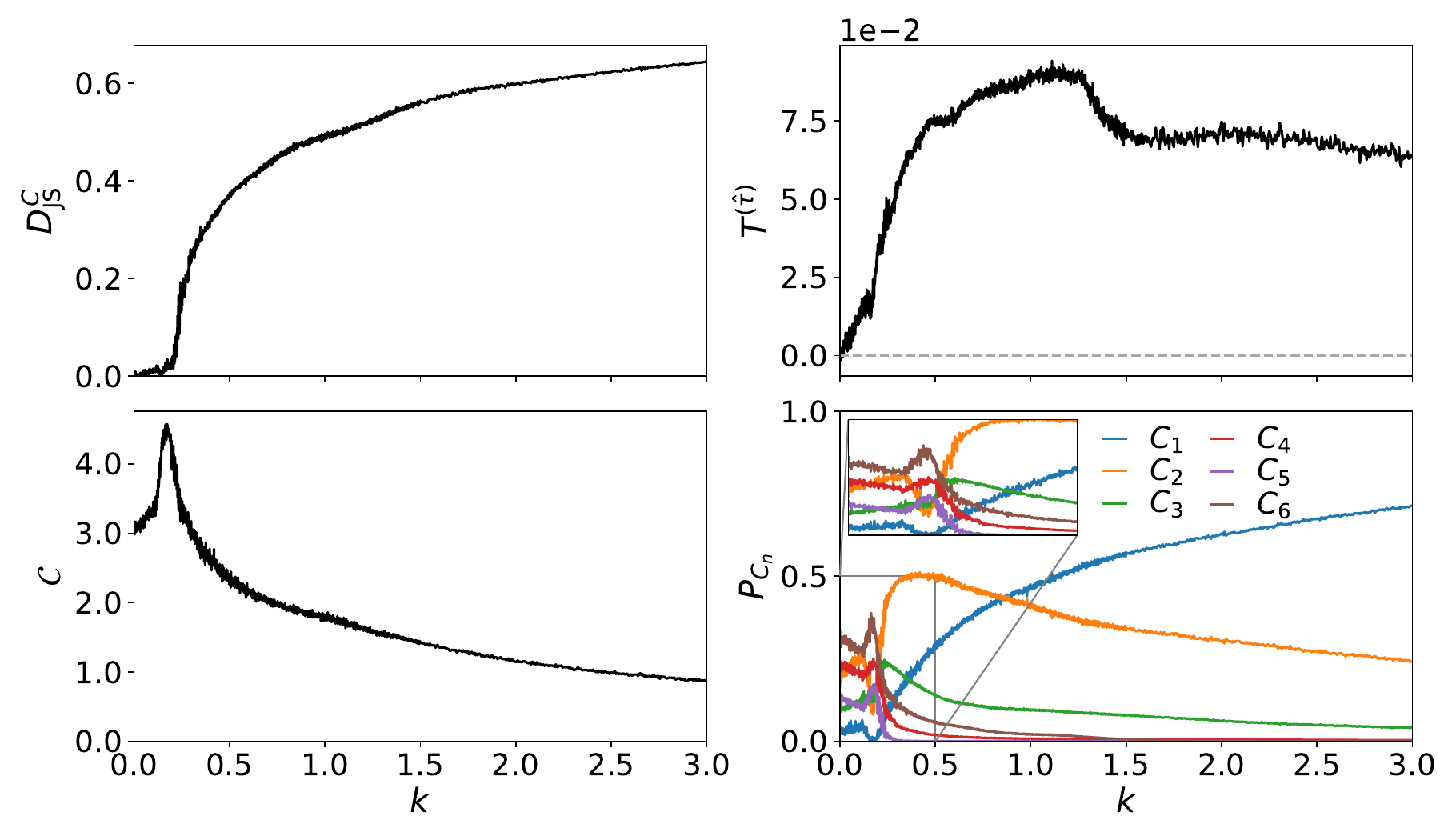} 
    \caption{Same as Fig.~\ref{fig:measureshenon} but for unidirectionally coupled Rössler oscillators. We generated time series of length $N=2^{18}$ of the $x$-components and derived ordinal patterns using an embedding dimension $d=6$ and an embedding delay $m=144$.}
    \label{fig:measuresroessler}
\end{figure}
Figure~\ref{fig:measuresroessler} summarizes our findings for the coupled R\"ossler oscillators.
$\djsc$ indicates weak to no interactions for $k\in[0,0.2]$. 
For higher coupling strengths, we observe a sudden increase in estimated strength of interaction, occurring at a coupling strength for which the \revise{oscillators} are expected to transition to phase synchronization. 
For even higher coupling strengths, we observe a monotonous increase of the estimated strengths of interaction.\\ 
$\tmi$ indicates directionality from
\revise{the first to the second oscillator}
throughout the whole range of coupling strengths, increasing monotonically with increasing coupling strength and showing a small decrease just before the transition to phase synchronization. \\
For low to intermediate coupling strengths, coupling complexity plateaus at values $\CC \approx 3$. It shows an increase for $k>0.1$, with a global maximum occurring just before the transition to phase synchronization. For higher coupling strengths, $\CC$ decreases monotonically as the 
\revise{individual dynamics}
become more similar. \\
As for the \revise{coupled Hénon maps}, higher order classes dominate the low to intermediate 
\revise{coupling} strengths $k\in[0,0.2]$, with an increase in the probability densities for higher order classes just before the transition, backing the findings obtained with $\CC$. For coupling strengths $k>0.2$, the higher order classes decrease, with $C_2$ becoming the predominant order class. We also observe a monotonous increase of $P_{C_1}$ for $k>0.2$ surpassing $P_{C_2}$ for $k\gtrsim1$.  \\

Our findings obtained for the coupled model systems show that $\djsc$ seems to underestimate the actual interaction strength for low coupling strengths. 
We also observed that all of the investigated estimators are susceptible to transitions between different synchronization regimes. 
$\tmi$, while yielding erroneous results for the coupled \revise{Hénon maps} at low coupling strengths, indicates the correct directionality for the \revise{coupled Rössler oscillators} throughout the investigated range of coupling strengths.
It increases steadily up until $k>1$, for which it starts to decrease again. \\
Otherwise, the \revise{transcript-based estimators for different properties of an interaction}
yielded results that match our expectations, except for very low coupling strengths, but nevertheless confirm previous findings~\cite{Monetti2009,Monetti2013CC,Monetti2013TMI,amigo2012} to a large extent. Additionally, coupling complexity and probability densities of order classes have shown to be sensitive to characteristics of dynamical behavior that could not be observed with $\djsc$ and $T^{\htau}$.
\revise{Before demonstrating the suitability of the transcript-based estimators for real-world applications,
we would like to refer to a recent study~\cite{adams2025} that investigated the robustness of estimators against various noise contaminations.}

\revise{
\section{Characterizing brain-wide interactions during different vigilance states}
}
In the following, we demonstrate how transcript-based estimators for different properties of an interaction can help to advance our understanding on how the brain's intrinsic activity is coordinated across space and time as a function of physiological states.   
To this end, we investigate continuous, multi-day electroencephalographic (EEG) recordings from nine subjects (3 females, age 19--81 years) that participated in earlier studies~\cite{Lehnertz2021a,Broehl2023a}. 
Four subjects demonstrated with and five subjects without disorders of the central nervous system
(CNS), however, none of the subjects exhibited any kind of pathological phenomena during the recording time.
All subjects signed informed consent that their data could be used and published for research purposes after being provided with written information and being given the opportunity to ask further questions. 
The studies were approved by the ethics committee of the University of Bonn and were conducted adhering to the principles outlined in the Declaration of Helsinki.

EEG data were continuously recorded for 2 to 7 days from 19 electrodes placed according to the International 10-20 system~\cite{Seeck2017}, and \revise{electrode} Cz served as physical reference.
Data were sampled at \unit[256]{Hz} using a 16 bit analog-to-digital converter and bandpass filtered offline between \unit[1~-~45]{Hz} (4th order Butterworth characteristic). 
A notch filter (3rd order) was used to suppress contributions at the line frequency (\unit[50]{Hz}). 
We visually inspected all recordings for strong artifacts (e.g. subject movements or amplifier saturation) and excluded such data from further analyses.

Taking into account the brain's nonstationarity~\cite{lehnertz2017}, we performed a moving-window analysis (nonoverlapping windows with a duration of \unit[20]{s} corresponding to $N=5120$ data points) and calculated --~for each window~-- all transcript-based estimators (with $d = 6$ and $m=1$) between all pairs of sampled brain regions.\\

\subsection{Temporal evolution of estimators}
\begin{figure}[htbp] 	
  \includegraphics[width=\linewidth]{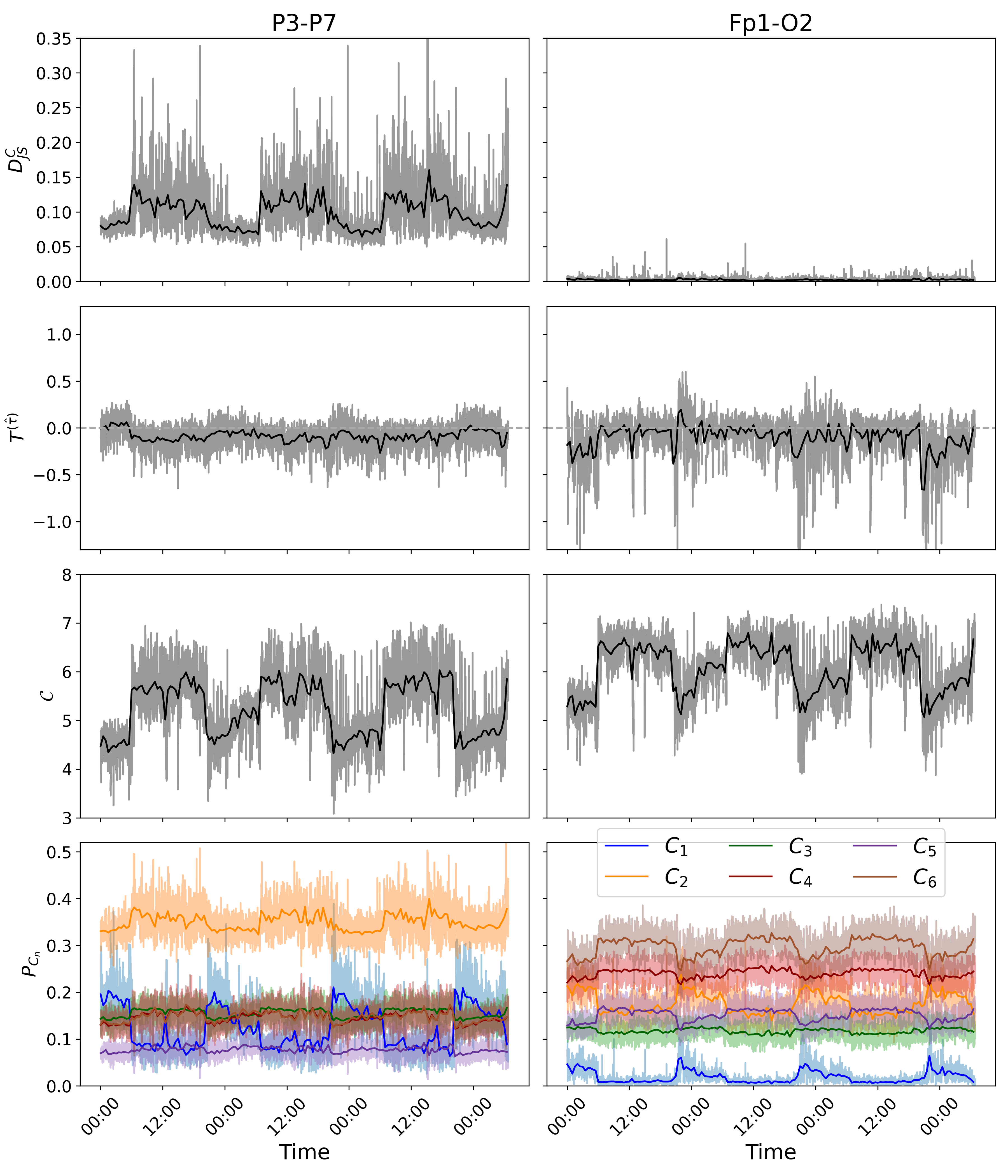}
 	\caption{ (From top to bottom) Time-evolution of estimated interaction strength, direction and complexity 
    for short-range (left) and long-range (right) interactions in a human brain. Dark lines and shaded areas indicate mean and standard deviation over successive $30$\,min time windows. \revise{We derived ordinal patterns using an embedding dimension $d=6$ and an embedding delay $m=1$.}}
 	\label{fig:timeev}
 \end{figure}

We first investigate the temporal evolution of transcript-based estimators for strength, direction, and complexity of an interaction over the whole EEG recording of a single subject (Fig. \ref{fig:timeev}), comparing a short-range interaction between neighboring pairs of brain regions (\revise{left precuneus, sampled with electrode P3 and left inferior temporal gyrus, sampled with electrode P7}) to a long-range interaction spanning over the whole brain (\revise{from the left superior frontal gyrus, sampled with electrode Fp1 to the right middle occipital gyrus, sampled with electrode O2}).\\
For the short-range interaction, $\djsc$ indicates a strength of interaction of $\djsc\approx0.1$ at nighttime, with a noticeable increase up to values of $\djsc\approx0.13$ during the day, for which fluctuations also greatly increase. In comparison, for the long-range interaction, $\djsc$ indicates almost no interaction at all, remaining mostly 0, save for some fluctuations.\\
For both interactions, $\tmi$ exhibits a tendency towards negative values.
While for the short-range interactions, the directionality index remains confined to a range around 0, for the long-range interactions we observe large fluctuations coinciding with the transition from \revise{daytime} to \revise{nighttime}. \\
For the short-range interactions, coupling complexity varies around $\CC \approx 5$ at nighttime, increasing up to $\CC \approx6$ at daytime, with transitions between 
\revise{night- and daytime}
clearly resolvable. 
For the long-range interactions, $\CC$ varies around $5$-$6$ at night- and $\approx 7$ at daytime, exhibiting smaller fluctuations than those seen for the short-range interactions. 
Comparing the differences in $\djsc$ between short- and long-range interactions to the only weakly pronounced differences in $\CC$ suggests that complexity estimators appear to enable the characterization of aspects of the dynamics, for which estimators of strength and direction are less sensitive.\\
The probabilities of order classes indicate a generally lower complexity 
for the short-range interactions, with $P_{C_2}$ dominating throughout the whole time range.
We also note that probability densities of order class $C_1$ and $C_2$ exhibit the highest sensitivity to the transition between \revise{night- and daytime},
with the probability densities of $C_{\geq3}$ differing only slightly between the two 
regimes. Additionally, almost no distinction can be made between $P_{C_3}$, $P_{C_4}$ and $P_{C_6}$. For the long-range interactions, we observe higher probability densities for higher order classes, with $P_{C_1}$ fluctuating around 0.5 and becoming close to 0 at daytime.
Additionally, for the long-range interactions, $P_{C_1}$ ($\CC$) attains its highest (lowest) values just after the transition to \revise{nighttime},
increasing (decreasing) steadily as 
\revise{night}  
This observation could be made to lesser extent also for the short-range interactions and indicates \revise{both} the sensitivity of complexity estimators to 
dynamical regimes \revise{and} the differences in information content that can be inferred from interactions between different pairs of brain regions.

\subsection{Spatial brain-wide distributions of interaction estimators}
For each EEG recording, we divided the data into two time ranges, namely daytime (6:00-22:00) and nighttime (22:00-6:00).
\revise{Lacking detailed somnographic evaluation but taking into account the clinically prescribed daily routine, we assumed subjects to be awake resp. asleep during these time ranges.
We calculated, for each time range, 
the averaged value of each interaction estimator 
from each recording site $c$ to the
remaining $N_c - 1$ sites as
\begin{equation}\label{eq:sumrec}
    \overline{\epsilon(c)} = \frac{1}{N_c-1} \sum_{c\neq c'} \epsilon(c’),
\end{equation}
where $\epsilon$ is a placeholder for $\djsc$, $\tmi$, $\CC$, and $P_{C_i}$ ($i\in [1,6]$). 
}
\revise{
\begin{figure}
    \centering
    \includegraphics[width=\linewidth]{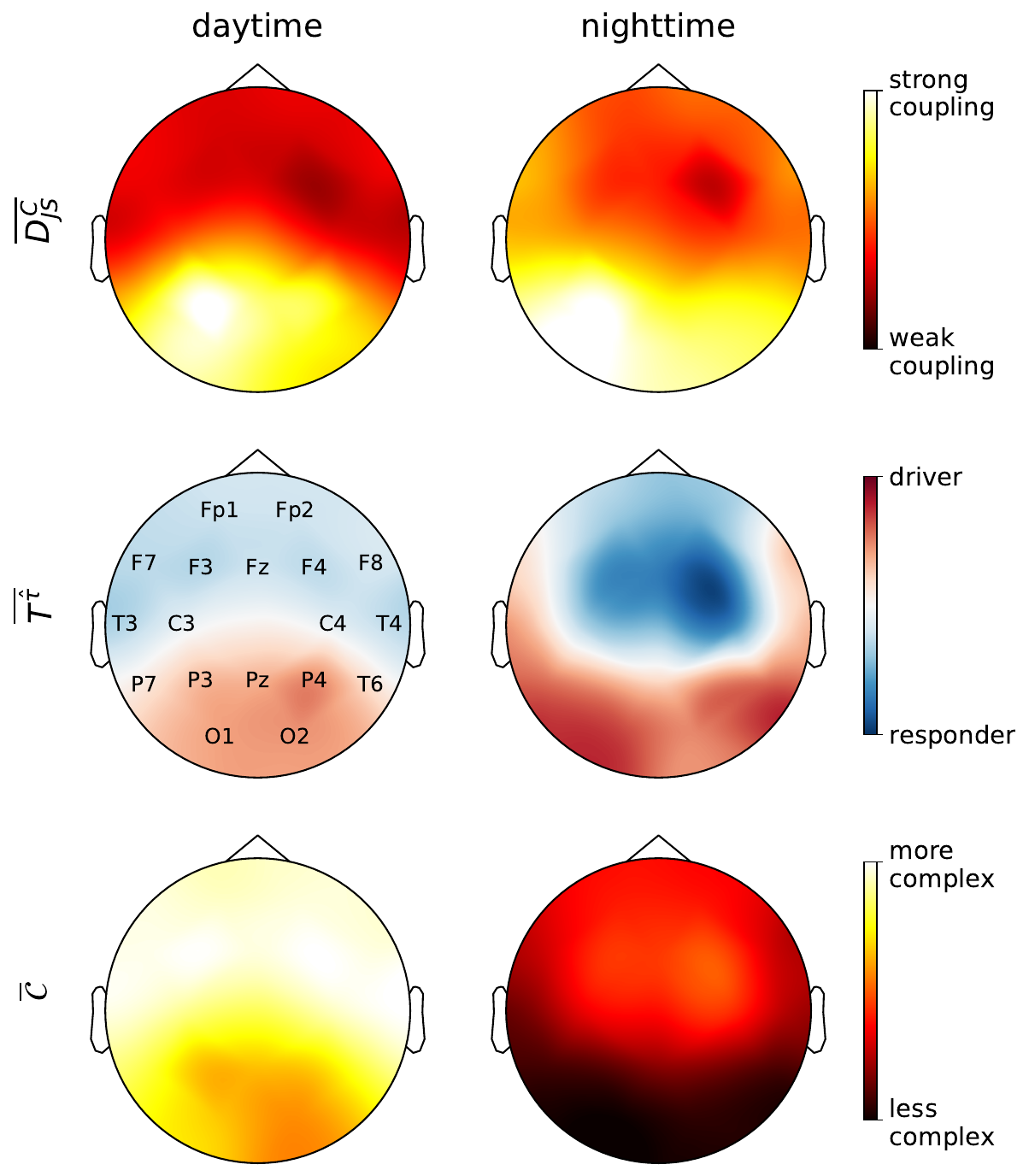}
    \caption{\revise{
        Spatial distributions of transcript-based estimators for strength (top), preferred direction of information flow (middle) and complexity $\CC$ (bottom) of brain-wide interactions during day- and nighttime projected onto the surface of the head (cubic polynomial interpolation).
        Grand averages over all subjects of the respective temporal means of mean estimators from each recording site to the remaining sites (cf. Eq. \ref{eq:sumrec}).
        Middle left plot highlights electrode positions and labels.}
        }
    \label{fig:djsccc}
\end{figure}
}
\revise{In Fig.~\ref{fig:djsccc} we show, as a grand average over all subjects, spatial distributions of estimators for strength, direction, and complexity of interaction.

For both day- and nighttime data, we observe strongest interactions to be spatially confined to the posterior part of the brain, indicative of a high abundance of short-range interactions
These are slightly more emphasized over the left brain hemisphere.
This spatial interaction structure most likely reflects interactions of the visual network and of the posterior part of the default mode network~\cite{doucet2011,gillig2025,lehnertz2025}.
During nighttime, we observe slightly stronger interactions of the temporal lobes, possibly reflecting processes related to memory consolidation during sleep~\cite{kaefer2022,lehnertz2025}.

We find a similar spatial patterning for the preferred direction of interaction, whereby the posterior regions can be identified as drivers of the anterior regions.
For the daytime data, this is consistent with earlier findings~\cite{martini2011,coito2019,das2021}. 
Interestingly, during nighttime we observe the temporal lobes to switch roles, now acting as driver, which possibly highlights their role in memory consolidation during sleep.

Coupling complexity highlights the most complex interactions during daytimes to be confined to the left and right fronto-temporo-central brain areas, evidently in those regions with only weak interactions but which subserve higher executive functions such as emotional regulation, planning, reasoning and problem solving~\cite{stuss2013}. As expected, such functions are reduced or absent during sleep, which is reflected in the overall reduced coupling complexity.\\
\begin{figure}[htbp]
    \centering    
    \includegraphics[width=\linewidth]{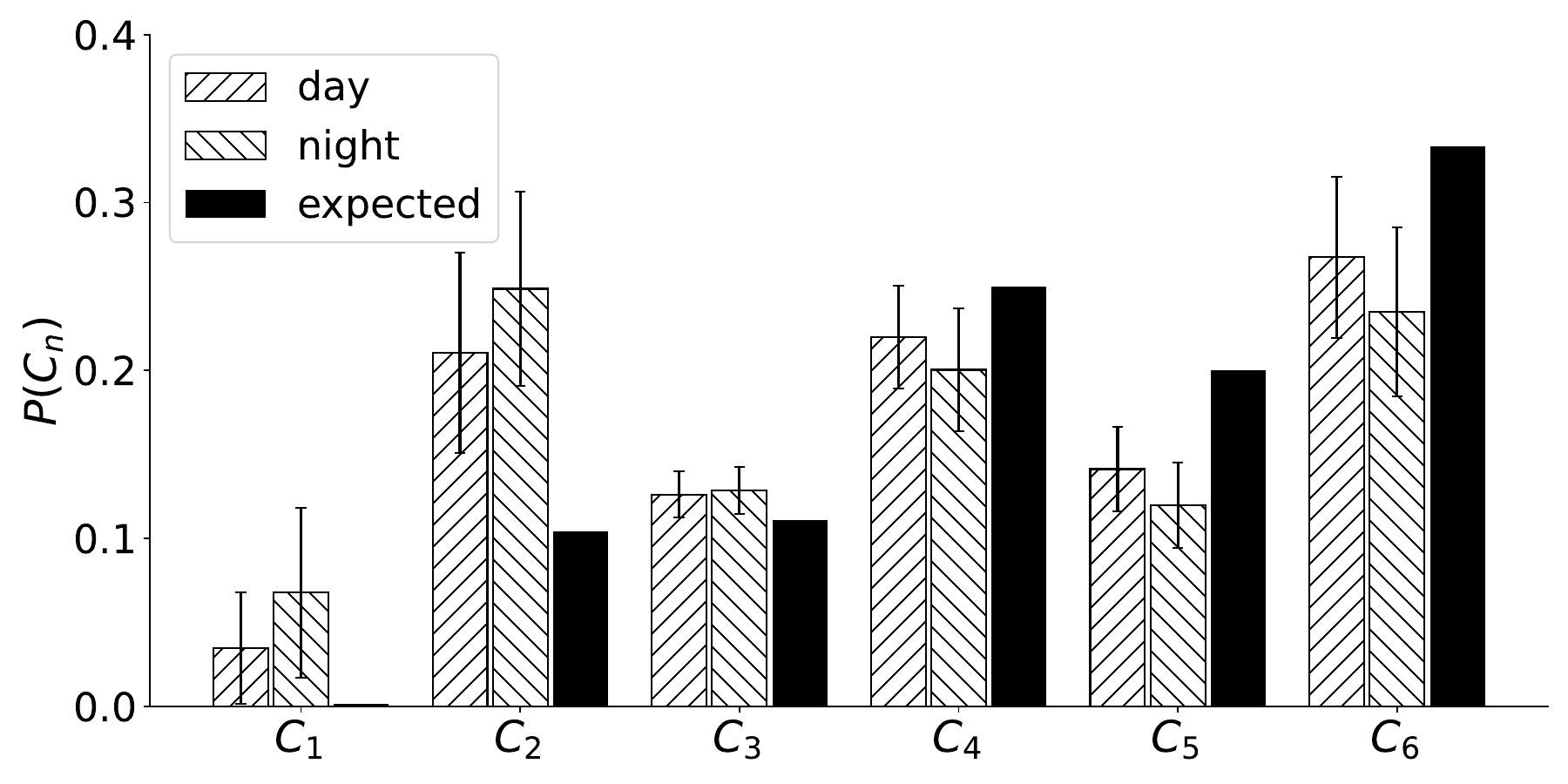} 
    \caption{\revise{Grand averages of probability densities $P$ of order classes $C_1$ -- $C_6$ estimated from daytime and nighttime data. Error bars reflect standard deviation over all subjects and all recording sites.
    Expected values for the respective densities were derived analytically~\cite{Wilf1986}.}
    }
    \label{fig:barplotoc}
\end{figure}

Before we proceed with the time-of-day-dependent spatial distributions of the complexity estimator that is based on probability densities of order classes, we briefly report an order-dependent impact on the densities: we observe the magnitudes of even-order densities to be generally larger than the odd-order ones.
This dependence can also be observed for 
the expected values for the respective probabilities of each order class (Fig.~\ref{fig:barplotoc}).}

\begin{figure}
    \centering
    \includegraphics[width=.97\linewidth]{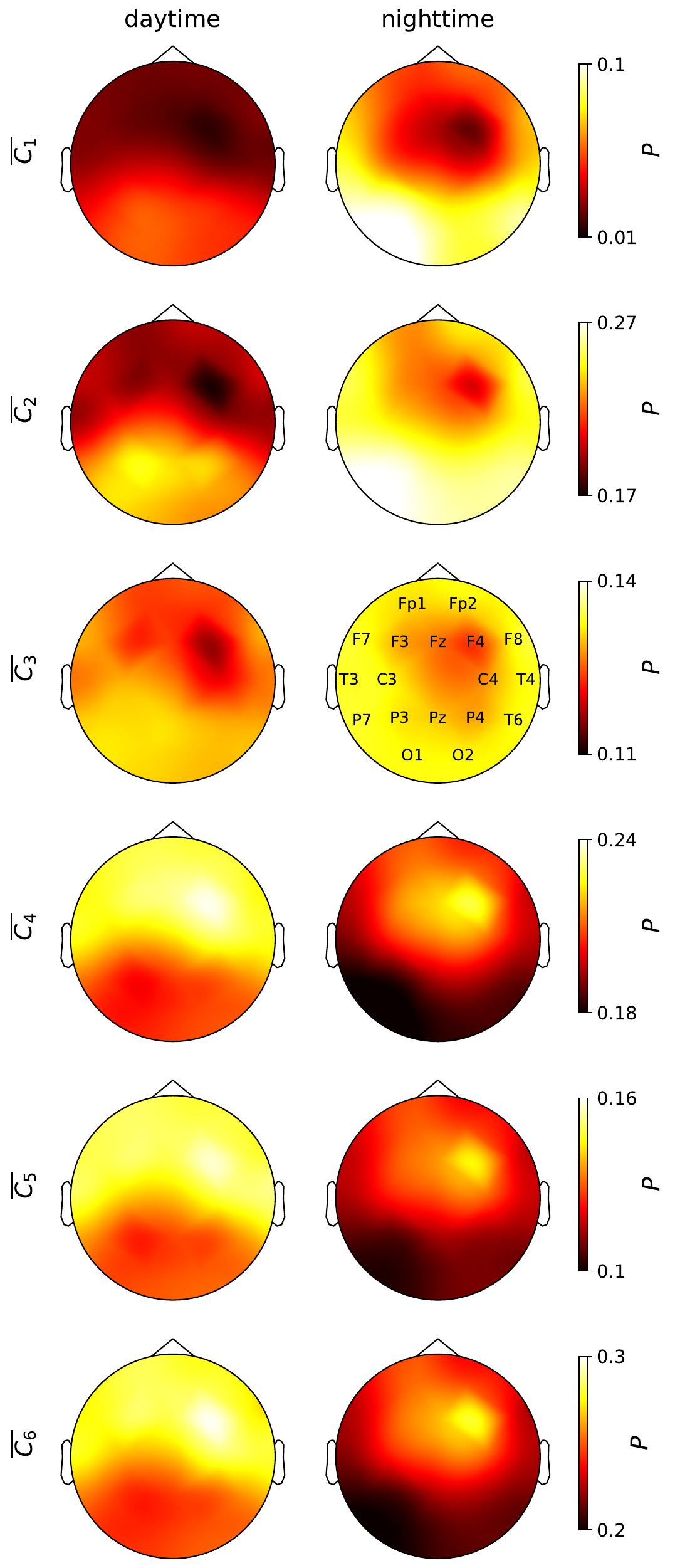}
    \caption{\revise{Same as Fig. \ref{fig:djsccc}, but for the probability densities of order classes $C_1$ to $C_6$ (from top to bottom).
    Note the different scalings of plots for each order class which resulted from the order-dependent impact on the densities (cf. Fig.~\ref{fig:barplotoc}).
    The third plot from the top on the right-hand side highlights electrode positions and labels.}}
    \label{fig:ocs}
\end{figure}

\revise{Figure~\ref{fig:ocs} summarizes our findings for interaction complexity based on 
the probability densities of order classes.
The mean probability densities of low order classes ($C_1$ and $C_2$) are highest during nighttimes, and their spatial distributions mostly highlight the posterior brain areas (visual network and posterior
part of the default mode network). This is to be expected, at least to a large extend, given the high similarity of sleep-related brain dynamics sampled from these areas, which is reflected in low permutation orders. In contrast, the complexity of brain dynamics and their interactions are highest during daytimes (awake or conscious~\cite{tononi1998} states), as indicated by the high mean probability densities of order classes $C_{\geq4}$ (order class $C_3$ takes an intermediate position). Their distributions highlight the left and right fronto-temporo-central brain areas, which subserve higher executive functions (cf. spatial distribution  of coupling complexity $\CC$ in Fig.~\ref{fig:djsccc}).
Interestingly, the mean probability densities of order classes $C_{\geq4}$ accentuate the right middle frontal gyrus (around electrode location F4), which exhibits a high interaction complexity during both day- and nighttimes. The diminished interaction complexity observed instead with low order classes $C_{\leq 3}$ may point to the relevance of this brain region, which is known to act as a gateway between top-down and bottom-up control of attention~\cite{corbetta2008}. 
Interaction complexity based on the probability densities of order classes thus appears to allow for a more fine-grained disentangling of state-dependent interactions between different 
brain regions, as compared to coupling complexity $\CC$.
}
%
\section{Conclusions}
\revise{We revisited the concept of transcripts, an extension of the concept of ordinal patterns that provides a means to characterize relations between two sequences of ordinal patterns via their algebraic relationship.
At the example of two paradigmatic models of coupled dynamical systems of varying complexity, we showcased the usage of transcript-based estimators for strength, direction, and complexity of an interaction, pointing out some limitations. Even though transcript-based estimators for strength and direction of interactions exhibited a low sensitivity for very weak couplings, our findings match the expectations we pose on estimators for strength and direction, in general~\cite{kreuz2007,osterhage2007,papana2011,porz2014,lehnertz-dickten2015,Krakovska2018,echegoyen2019,Lehnertz2023}.
A comparison of transcript-based estimators for the complexity of an interaction with other related estimators~\cite{haruna2019,haruna2023} has yet to be performed.
 
Our analyses of continuous, multi-day recordings of ongoing, i.e. non-task-related brain dynamics from nine subjects using the singular concept of a transcript-based characterization of interactions allowed us to confirm some prior findings but also to shed new insights into brain-wide interactions underlying different vigilance states.
Results obtained with estimators for strength and direction of interactions largely confirmed earlier characterization of spatial-temporal dynamics of the human brain~\cite{Lehnertz2023,boaretto2023}.
However, particularly the complexity estimator that is based on probability densities of order classes yielded valuable results, which may contribute to a better understanding of the complex interactions in the human brain when transiting between different physiological and pathophysiological states.

We are confident that transcript-based estimators for different properties of an interaction carry the potential to improve characterization of couplings between complex dynamical systems, and particularly of the still poorly understood spatial-temporal interaction dynamics in the human brain.
}

\begin{acknowledgments}
The author acknowledges fruitful discussions with Timo Br{\"o}hl and Max Potratzki. 
MA and KL acknowledge support from the Verein zur Foerderung der Epilepsieforschung e.V. (Bonn).
\end{acknowledgments}


\section*{AUTHOR DECLARATIONS}

\subsection*{Conflict of Interest}
The authors have no conflicts of interest to disclose.

\subsection*{Data availability}
The EEG datasets presented in this article are not readily available because they contain information that could compromise the privacy of research participants. Requests to access the datasets should be directed to the corresponding author. 
The code that was used to analyze the data in this study is openly available on 
Github: \\
\url{https://github.com/Mannystein/Transcripts/}


\appendix

\section{Theoretical maximum for $\djsc=\infdivD{P_C}{P_C^{\textnormal{ind}}}$}\label{app:maxdjsc}
\begin{figure}[]
    \centering
    \includegraphics[width=\linewidth]{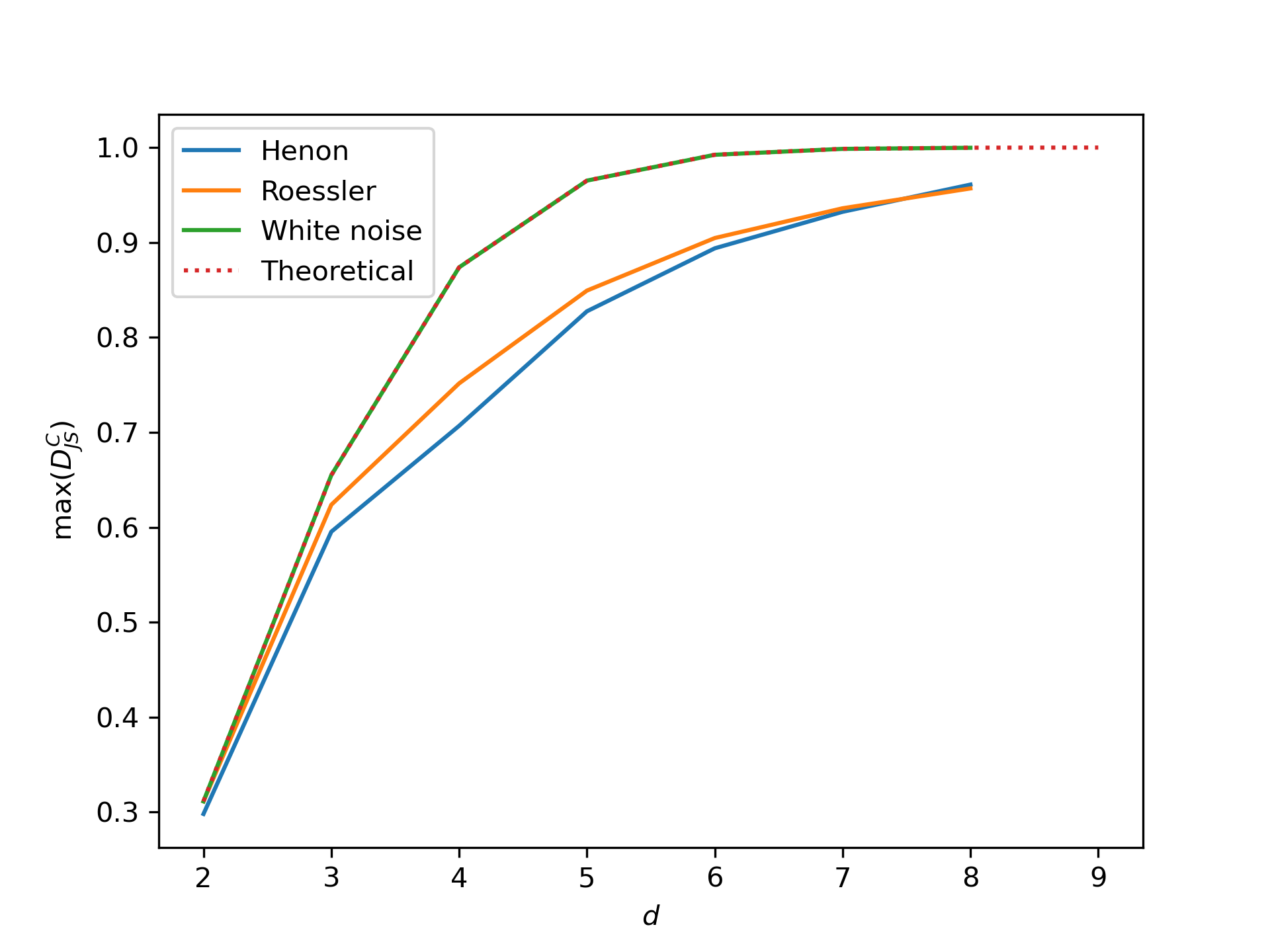}
    \caption{Dependence of the maximum obtained value for $\djsc$ on the embedding dimension for different dynamical systems investigated in this paper. The existence of forbidden patterns leads to a deviation from the theoretical maximum.}
    \label{fig:maxdjsc}
\end{figure}

The theoretical maximum of $\djsc$ (Eq. (\ref{eq:djsc})) is obtained, if the probability densities $P_{C_n}$ and  $P^\text{ind}_{C_n}$ are orthogonal. In this ideal case, we obtain $\djsc=\log_2(2)=1$.
However, for probability spaces with a small number of possible states, this maximum cannot be reached in practice. Specifically, the limited number of states constrains the degree of distinguishability between the joint and independent distributions, preventing complete orthogonality. As the embedding dimension and thus also the number of possible transcripts and order classes is increased, the discretization of the probability densities becomes increasingly fine-grained and the maximum possible value of $\djsc$ converges towards its theoretical upper bound of 1. In order to provide comparability of the acquired results, we here assess the dependence of the maximum Jensen-Shannon divergence on the embedding dimension.
The probability densities of order classes are distributed based on the relative number of ordinal pattern that belong to a certain order class.
The maximum attainable value for $\djsc$ is reached for two identical time series, for which the probability densities of ordinal patterns from both time series are uniformly distributed.
The joint probability densities of order classes reduce to
\begin{equation*}
    P_{C_1} = 1, \quad P_{C_{>1}} = 0.
\end{equation*}
For the independent probabilities, we obtain 
\begin{equation*}
    P^\text{ind}_{C_1} = \frac{1}{d!} \quad \text{and} \quad  \sum_{n=1}^{M} P_{C_n}= 1.
\end{equation*}
In this case, the Jensen-Shannon divergence for order classes becomes
\begin{equation*}
    \begin{aligned} 
    \infdivD{P_C}{P_C^\text{ind}} &= -\frac{1+P^\text{ind}_{C_1}}{2}\log_2\left(\frac{1+P^\text{ind}_{C_1}}{2}\right) \\
    &\qquad + \frac{1}{2}(1-P^\text{ind}_{C_1})+\frac{P^\text{ind}_{C_1}}{2}\log_2(P^\text{ind}_{C_1}) \\
    &= -\frac{d!+1}{2d!}\log_2\left(\frac{d!+1}{2d!}\right) \\
    &\qquad + \frac{d!-1}{2d!}+\frac{1}{2d!}\log_2(d!).
    \end{aligned}
\end{equation*}
It follows that $\infdivD{P_C}{P_C^{ind}}\rightarrow 1$ as $d \rightarrow \infty$ $(P_{C_1}\rightarrow 0)$. \\
In practice and for a given embedding dimension, the number of ordinal patterns in a given order class can be obtained by calculating the order of every ordinal pattern of length $d$ and counting. 
The mixing probability $P_{C_n}^{\text{M}}$ can then be calculated as in Eq.~(\ref{eq:pm}). 
We can then use Eqs.~(\ref{eq:ekl}) and (\ref{eq:djsc}) to obtain the Jensen-Shannon divergence. The obtained independent probabilities as well as the resulting maximum Jensen-Shannon divergence are listed in Table~\ref{tab:maxdjsc}. 
It should be noted that, due to forbidden patterns~\cite{Amigo2008}, altering the ordinal patterns' distributions, the theoretical maximum is, in general, not reached for a dynamical system whose state space differs from that of white noise. Fig.~\ref{fig:maxdjsc} shows the maximum Jensen-Shannon divergence obtained for different model systems compared to the theoretically possible maximum value.

\begin{table}[h!]
\begin{tabular}{l|llllll|l}
 d   & $C_1$ & $C_2$ & $C_3$ & $C_4$ & $C_5$ & $C_6$ & $\max{(\djsc)}$ \\ \hline
$2$ &  1/2  &  1/2  &    &    &    &  & 0.311  \\ \hline
$3$ &  1/6  &  3/6  &  2/6  &    &    & & 0.655    \\ \hline
$4$ &  1/24  &  9/24  &  8/24  &  6/24  &    &  & 0.874  \\ \hline
$5$ &  1/120  &  25/120  &  20/120  &  30/120  &  24/120  &  20/120 & 0.965  \\ \hline
$6$ &  1/720  &  75/720  &  80/720  &  180/720  &  144/720  &  240/720 & 0.992 \\ \hline
\end{tabular}
\caption{Independent probabilities of order classes and maximum attainable value for $\djsc$ for different embedding dimensions.}
\label{tab:maxdjsc}
\end{table}

\revise{
\section{Bounds for $\tmi$}\label{app:boundtmi}
The upper bound of the mutual information between two random variables $A$ and $B$ can be derived as 
\begin{equation}
    \max{(I(A,B))} = \min(H(A),H(B))) = \log\left[\min(|\A|,|\B|)\right],
\end{equation}
with $|\A|$ and $|\B|$ indicating the cardinality of the alphabets $\A$ and $\B$ of the respective random variables. Through the process of creating the ordinal pattern sequences $\hA$ ($\hB$) from observations of $A$ ($B$), the upper bound of the mutual information translates to 
In our case, both sets of possible states are $S_d$, whose cardinality is $d!$. Thus, the upper bound for the mutual information between two symbol series $\halpha$ and $\hbeta$ reduces to 
\begin{equation}
    \max{(I(\hA,\hB))} = \log d!,
\end{equation}
and thus 
\begin{equation}
    -\log d! \leq \tmi \leq \log d!.
\end{equation}
}

%

\end{document}